\documentclass{sigchi-ext}
\usepackage[T1]{fontenc}
\usepackage{textcomp}
\usepackage[scaled=.92]{helvet} 
\usepackage{graphicx} 
\usepackage{balance}  
\usepackage{booktabs} 
\usepackage{ccicons}  
\usepackage{ragged2e} 

\usepackage{paralist}
\usepackage{xcolor}
\usepackage{etoolbox}
\usepackage{hyperref}
\usepackage{multirow}
\usepackage{tabularx}
\def\plaintitle{SIGCHI Extended Abstracts Sample File: Note Initial
  Caps} 
\def\emptyauthor{}
\def\plainkeywords{Transparency; Explainability; Interpretability}

\title{A Multistakeholder Approach Towards Evaluating AI Transparency Mechanisms}

\numberofauthors{7}
\author{%
  \alignauthor{%
    \textbf{Ana Lucic}\\
     \affaddr{Partnership on AI} \\ 
    \affaddr{University of Amsterdam} \\
    \affaddr{Amsterdam, Netherlands} \\
    \affaddr{ana@partnershiponai.org} \\ \medskip
    \textbf{Madhulika Srikumar}\\
    \affaddr{Partnership on AI} \\
    \affaddr{San Francisco, USA} \\
    \email{madhu@partnershiponai.org} }\alignauthor{%
    \textbf{Ankur Taly}\\
    \affaddr{Google}\\
    \affaddr{San Francisco, USA}\\
    \email{ataly@google.com} \\ \medskip
    \textbf{Q. Vera Liao}\\
    \affaddr{IBM Research AI}\\
    \affaddr{New York City, USA}\\
    \email{vera.liao@ibm.com} } \vfil \alignauthor{%
    \textbf{Umang Bhatt}\\
    \affaddr{University of Cambridge}\\
    \affaddr{Mozilla Foundation}\\
    \affaddr{Cambridge, UK}\\
    \email{usb20@cam.ac.uk} \\ \medskip
    \textbf{Alice Xiang}\\    
    \affaddr{Sony AI}\\
    \affaddr{Tokyo, Japan}\\
    \email{alice.xiang@sony.com} }\alignauthor{%
    \textbf{Maarten de Rijke}\\
    \affaddr{University of Amsterdam}\\
    \affaddr{Ahold Delhaize Research}\\
    \affaddr{Amsterdam, Netherlands}\\
    \email{m.derijke@uva.nl} } }

\definecolor{linkColor}{RGB}{6,125,233}
\hypersetup{%
  pdftitle={\plaintitle},
  pdfauthor={\emptyauthor},
  pdfkeywords={\plainkeywords},
  bookmarksnumbered,
  pdfstartview={FitH},
  colorlinks,
  citecolor=black,
  filecolor=black,
  linkcolor=black,
  urlcolor=linkColor,
  breaklinks=true,
}

\begin{document}

\CopyrightYear{2021}
\setcopyright{rightsretained}
\conferenceinfo{CHI'21}{May 8-13, 2021, Online Virtual Conference }
\isbn{978-1-4503-6819-3/20/04}
\doi{https://doi.org/10.1145/3334480.XXXXXXX}
\copyrightinfo{\acmcopyright}

\maketitle

\RaggedRight{} 

\begin{abstract}
Given that there are a variety of stakeholders involved in, and affected by, decisions from machine learning (ML) models, it is important to consider that different stakeholders have different transparency needs \cite{mohseni2020multidisciplinary}.
Previous work found that the majority of deployed transparency mechanisms primarily serve technical stakeholders \cite{bhatt_explainable_2019}. 
In our work, we want to investigate how well transparency mechanisms might work in practice for a more diverse set of stakeholders by conducting a large-scale, mixed-methods user study across a range of organizations, within a particular industry such as health care, criminal justice, or content moderation. 
In this paper, we outline the setup for our study.

\end{abstract}





\printccsdesc

\section{Introduction}
\label{section:introduction}

There is an increased demand for transparency in artificial intelligence (AI) systems. 
So far, the explainable AI (XAI) community has primarily contributed computational methods for understanding predictions of machine learning (ML) models~\cite{guidotti-2018-survey}. Such methods help users understand the rationale behind a model's behavior and
typically range from global techniques that explain the entire model to local techniques that explain predictions from individual instances~\cite{guidotti-2018-survey}. 


Previous work has shown that there exists a significant gap between research and deployment of transparency mechanisms for ML models: although many types of stakeholders are involved in the deployment of ML models, scenarios where transparency mechanisms are currently deployed are almost exclusively for stakeholders who build, validate, or deploy ML models \cite{bhatt_explainable_2019}. 
This may be because most transparency mechanisms come in the form of local explanations, and although this is where the XAI community has focused a large portion of its efforts \cite{guidotti-2018-survey}, perhaps it is not necessarily what stakeholders need in practice~\cite{bhatt2020machine}, other than ML engineers.

We argue that in order to meet the needs of non-technical stakeholders, we should think more broadly in terms of what transparency mechanisms are offered, going beyond explaining a model's \emph{behavior} under certain conditions (i.e., its predictions), by also offering the \emph{process} that went into building the model as an explanation. 
For example, a Model Card \cite{mitchell_model_2019} might be more useful for an executive who is making a decision about whether or not to deploy a model at scale across their organization compared to seeing SHAP values \cite{lundberg_unified_2017} for five particular input instances.

To examine the effectiveness of various transparency mechanisms, we are building off of the previous work \cite{bhatt_explainable_2019} conducted at our organization, the Partnership on AI: a global multistakeholder non-profit organization that aims to develop and share best practices for responsible use of AI.\footnote{\url{https://www.partnershiponai.org}} 
This work was focused on taking stock of how transparency mechanisms are currently deployed across a range of industries and organizations. 
In our current work, we want to investigate how well these mechanisms work in practice by conducting a large-scale, mixed-methods, multistakeholder study on 
\begin{inparaenum}[(a)]
    \item providing meaningful explanations relevant to the specific needs of diverse stakeholders in different use cases, and 
    \item determining how these explanations should be evaluated. 
\end{inparaenum}
We want to focus on a particular industry (i.e., content moderation, criminal justice, health care, etc) to enable application-grounded evaluation \cite{doshi-2017-towards}, which involves real users conducting real tasks. 
We plan to examine the following research questions:

\textbf{RQ1:} What types of transparency mechanisms are most appropriate for different stakeholders in different use cases? 

\textbf{RQ2:} How can we evaluate different types of transparency mechanisms in (a) objective terms such as a user's ability to perform a task using an explanation, and (b) subjective terms such as the impact on a user's trust in an AI system?

\section{Preliminaries}
A fundamental problem in the existing XAI literature is that the notions of transparency, interpretability and explainability are not well-defined and are often used interchangeably \cite{kaur2020interpreting,lipton_mythos_2016}. 
We therefore explicitly define the terms used in our work as follows: 

\textbf{Transparency}: providing insight into an ML model. Two possible forms of transparency include: \textit{behavior-based transparency} and \textit{process-based transparency}.
    
    
\textbf{Behavior-based transparency}: providing insight into how an ML model makes decisions, in a global or local manner, from an algorithmic or mathematical perspective. This is also sometimes referred to as \textit{interpretability}. Some models are considered to be ``inherently interpretable'' (i.e., shallow decision trees or linear models with a small number of features), while others require \emph{post-hoc} methods to generate these interpretations (i.e., SHAP values \cite{lundberg_unified_2017}, LIME feature importances \cite{ribeiro-2016-should}, counterfactual examples \cite{lucic_actionable_2020,wachter_counterfactual_2017}). 

\textbf{Process-based transparency}: providing insight into the whole ML \emph{modeling pipeline}, from development to production (i.e., models' intended use, data provenance, data collection, data splits for training and evaluation, team responsible for development and monitoring, evaluation metrics, reporting and visualization, etc. \cite{gebru_datasheets_2020,mitchell_model_2019}). 


\textbf{Explainability}: translating transparency insights into something that is understandable to a human. 
Explainability may require either \textit{behavior-based} or \textit{process-based} transparency (or both), along with other information. 
It is conditioned on (a) the stakeholder's needs and characteristics, and (b) the use case in which it is deployed. For models that are considered to be ``inherently interpretable'' \cite{rudin2019stop}, this translation is still necessary since we need to decide how to present the information to the stakeholder.
	
\textbf{Stakeholder}: individuals who have a vested interest in the transparency of a system \cite{bhatt_explainable_2019}.
	
\textbf{Use case}: a particular context in which transparency is used or required.

Disentangling transparency and explainability in this way allows us to 
\begin{inparaenum}[(a)]
    \item present \textit{process-based} transparency information as a potential explanation, and 
    \item separate the algorithmic component of generating model insights (i.e., the \textit{interpretation}), from the form in which the information is presented to the user (i.e., the \textit{explanation}).
\end{inparaenum}
This allows us to have different explanations for different stakeholders, while using the same underlying information for transparency. 

For example, in the context of \textit{behavior-based} transparency, an end user might only be interested in a ranking of the most important features (e.g., ordered set of features based on mean SHAP \cite{lundberg_unified_2017} values), while an ML engineer might need more granular information (e.g., plots with individual SHAP values \cite{lundberg_unified_2017} for every sample, where each feature is shown on a separate plot). 
Although the underlying \textit{interpretability} mechanism is the same (i.e., SHAP \cite{lundberg_unified_2017}), the resulting \textit{explanations} are different, and can therefore be tailored to the stakeholders' needs. 


\section{Stakeholder-Informed Study Design}
\label{section:participatory}
To construct our study, we plan to solicit input from relevant stakeholders in order to ensure the study represents tasks and subjective questions that reflect their values and transparency needs. 
This could take various forms including review panels, group workshops, and/or individual interviews with stakeholders, in particular those who interact with transparency mechanisms, in order to:
\begin{inparaenum}[(a)]
    \item uncover common themes in participants' encounters and experiences with transparency techniques, and identify which type of explanations would be best suited for their use case, and
    \item construct scenario-style sessions modeling real use cases to examine participants' transparency needs \cite{brown2020_public}
\end{inparaenum}


We will first design a pilot study, where stakeholders interact with various types of existing transparency mechanisms, both \textit{behavior-} and \textit{process-based}, and provide feedback on their experiences as they do so. 
The goal here would be to
\begin{inparaenum}[(a)]
    \item identify which type of transparency mechanism is best suited for each stakeholder's particular use case, and
    \item determine how best to translate the information provided by the underlying transparency mechanism into an explanation. 
\end{inparaenum}
These explanations would then be used as input for the user study outlined in the following section. 

\section{User Study}
\label{section:application}
Based on the input we receive from stakeholders, we will design an application-grounded evaluation \cite{doshi-2017-towards} study (i.e., a study with real users performing real tasks), in order to answer \textbf{RQ1}. 
This would involve having stakeholders perform a set of industry-specific objective tasks, as well as answer some subjective questions about their experiences (e.g., Likert-scale). 
Examples of such tasks include forward or counterfactual simulations \cite{doshi-2017-towards}. 
To elicit mental models \cite{hoffman_metrics_2018} of how the ML model works, we could encourage stakeholders to think aloud while performing the tasks. 


In order to test the effect on stakeholders' abilities to perform tasks, we would conduct a between-subject study, where all stakeholders are asked to perform the same tasks, but half would get some form of explanation while the other half would not. 
This would answer \textbf{RQ2a}. 

To answer \textbf{RQ2b}, we would include a within-subject study for the stakeholders who had explanations by asking the same set of subjective questions (i.e., the Trust Scale in \cite{hoffman_metrics_2018}) before and after completing the task, to see the effect that interacting with the explanations had on stakeholders' trust in, and satisfaction with, 
\begin{inparaenum}[(a)]
    \item the underlying model, and
    \item the explanation.
\end{inparaenum}



\section{Use Cases for Transparency}
\label{section:contexts}
In this section, we outline some examples of use cases we hope to elicit by interacting with stakeholders within a particular industry. 
We plan to use the HCXAI workshop to narrow our focus regarding the possible use cases, based on prior work such as~\cite{liao2020questioning}. 

\textbf{Interpreting individual predictions:}
providing an understanding of the salient factors for a particular prediction made by an ML model. 

\textbf{Gaining knowledge:}
generating new insights about the domain such as important decision factors or mechanisms \cite{liao2020questioning}, as well as understanding properties of the underlying dataset and task.




\textbf{Aiding decisions:}
offering supporting evidence for a prediction, which allows the decision-maker to choose how to incorporate this information with their own knowledge in order to make a decision.




\textbf{Suggesting interventions:}
suggesting appropriate interventions to the stakeholder in order to obtain a more favorable outcome, either from the model or in the real world. 




\textbf{Adapting system usage or control:}
allowing stakeholders to find the optimal ways to use the system, for example by adjusting their profiles or control settings.

\textbf{Model improvement:}
offering insights that enable stakeholders to improve the model.

\textbf{Model auditing:}
allowing investigation of concerns around model safety, ethics, and privacy.




\section{Conclusion}
Our work would be an important step in developing transparency mechanisms that are actually useful in practice to a diverse set of stakeholders. 
Model transparency is a multi-faceted problem, which does not have a single solution, and therefore the proposed solutions must be specific to both the use case and stakeholder involved. 
So far, we have had initial scoping conversations with over 15 partner organizations to gauge interest in the project, and are in the process of identifying an industry to center the study on. 
We are also designing a set of questions for soliciting input from stakeholders. 
We have also submitted a panel proposal to RightsCon\footnote{\url{https://www.rightscon.org/}} with the aim of facilitating a conversation between the XAI, HCI and human rights communities.

\pagebreak

\balance{} 

\bibliographystyle{SIGCHI-Reference-Format}
\bibliography{xai_bib}

\end{document}